# Efficient Wrapper/TAM Co-Optimization for SOC Using Rectangle Packing

Md. Rafiqul Islam, Muhammad Rezaul Karim, Abdullah Al Mahmud,
Md. Saiful Islam, Hafiz Md. Hasan Babu

Department of Computer Science and Engineering
University of Dhaka, Dhaka-1000, Bangladesh
rafik@udhaka.net ,{r_karimcs,aamrubel,sohel_csdu}@yahoo.com, hafizbabu@hotmail.com

**Abstract** : The testing time for a system-on-chip(SOC) largely depends on the design of test wrappers and the test access mechanism(TAM).Wrapper/TAM co-optimization is therefore necessary to minimize SOC testing time . In this paper, we propose an efficient algorithm to construct wrappers that reduce testing time for cores. We further propose a new approach for wrapper/TAM co-optimization based on two-dimensional rectangle packing. This approach considers the diagonal length of the rectangles to emphasize on both TAM widths required by a core and its corresponding testing time.

### 1.Introduction

Pre-designed and pre-verified intellectual property(IP) cores are being increasingly used in complex system-on-a-chip. However, testing these systems is difficult, and manufacturing test is widely recognized as a major bottleneck in SOC design. The general problem of SOC test integration includes the design of TAM architectures, optimization of the core wrappers, and test scheduling. Test wrappers form the interface between cores and test access mechanisms (TAMs), while TAMs transport test data between SOC pins and test wrappers [3]. We address the problem of designing test wrappers and TAMs to minimize SOC testing time. While optimized wrappers reduce test application times for the individual cores, optimized TAMs lead to more efficient test data transport on-chip. Since wrappers influence TAM design, and vice versa, a co-optimization strategy is needed to jointly optimize the wrappers and the TAM for an SOC.

In this paper we propose a new approach to integrated wrapper/TAM co-optimization and test scheduling based on a general version of rectangle packing considering diagonal length of the rectangles to be packed. The main advantages of our approach are that it minimizes the test application time as well as TAM utilization.

The rest of the paper is organized as follows. Related work is described in section 2 and a new approach to wrapper design is given in section 3. In Section 4,we formulate the wrapper/TAM co-optimization problem as a generalized version of rectangle packing. In section 5,we integrate the wrapper design algorithm with the test scheduling algorithm  to obtain an effective wrapper/TAM architecture and a test schedule that minimizes testing time. Finally, in Section 6, we present experimental results on one academic SOC.

### 2.Related work

Most prior research has either studied wrapper design and TAM optimization as independent problems, or not addressed the issue of sizing TAMs to minimize SOC testing time [1,9,10]. Alternative approaches that combine TAM design with test scheduling [2,8] do not address the problem of wrapper design and its relationship to TAM optimization.

The first integrated method for Wrapper/TAM co-optimization was proposed in [5,6,7].[5,7] are based on fixed-width TAMs  which are inflexible and result in inefficient usage of TAM wires. An approach to wrapper/TAM co-optimization based on a generalized version of rectangle packing was proposed in [6].This approach provides more flexible partitioning of the total TAM width among the cores.

### 3. Proposed Wrapper Design

The purpose of our wrapper design algorithm is to construct a set of wrapper chains at each core. A wrapper chain includes a set of the scanned elements (scan-chains, wrapper input cells and wrapper output cells).The test time at a core is given by:

$$T_{core} = p \times [1+\max\{s_i,s_o\}] + \min\{s_i,s_o\}$$

where p is the number of test vectors to apply to the core and si (so) denotes the number of scan cycles required to load (unload) a test vector (test response)[5]. So, to reduce test time , we should minimize the longest wrapper chain (internal or external or both), *i.e.* max{si, so}.Recent research on wrapper design has stressed the need for balanced wrapper scan chains [5,10] to minimize the longest wrapper chain. *Balanced* wrapper scan chains are those that are as equal in length to each other as possible.

The proposed Wrapper_Design algorithm tries to minimize core testing time as well as the TAM width required for the test wrapper. The objectives are achieved by balancing the lengths of the wrapper scan chains and imposing an upper bound on the total number of scanned elements.

Our heuristic can be divided in two main parts; the first one for combinational cores and the second one for sequential cores. For combinational cores, there are two possibilities. If *I+O*(where *I* is the number of functional inputs and **O** the number of functional outputs) is below or equal to the TAM bandwidth limit, $W_{max}$, then nothing is done and the number of connections to the TAM is **I+O**. If **I+O** is above $W_{max}$, then some of the cells on the I/Os are chained together in order to reduce the number of needed connections to the TAM.

---

procedure Wrapper_Design (int $W_{max}$, Core C)
{
　　//$W_{max}$ =TAM width
Total_Scan_Element= total IO+
$\sum$ C.Scan_Chain_Length[i]($1 \leq i \leq$ #SC);
1. If C.#SC=0
　　If ( Total_Scan_Element $\leq W_{max}$ )
　　　Assign one bit on every I/O wrapper cell;
　　Else
　　　Design $W_{max}$ wrapper scan chains;
2. Else
　　Mid_Lines = $W_{max}$ / 2;
　　Peak_Scan_Element = Total_Scan_Element / Mid_Lines ;
　　Sort the internal scan chains in descending order of their length;
　　For each scan chain SC
　　　For each wrapper scan chain W already created
　　　　If ( Length(W)+Length(SC) $\leq$ Peak_Scan_Element )
　　　　　Assign the scan chain to this wrapper scan chain W ;
　　　　Else
　　　　　Create a new Wrapper scan chain $W_{new}$ ;
　　　　　Assign the scan chain to this wrapper scan chain $W_{new}$ ;
Add functional I/O to balance the wrapper chains ;
}

---

Figure 1: algorithm for wrapper design

For sequential cores, at first an upper bound is specified(Peak_Scan_Element). The internal scan chains are then sorted in descending order. After that, Each internal scan chain is successively assigned to the wrapper scan chain, whose length after this assignment is closest to, but not exceeding the length of the upper bound. In our algorithm, a new wrapper scan chain is created only when it is not possible to fit an internal scan chain into one of the existing wrapper scan chains without exceeding the length of the upper bound. At last, functional inputs and outputs are added to balance the wrapper scan chains.

Our wrapper design algorithm gives results like table 1. Unlike [5], our Pareto-optimal points and their corresponding TAM utilized values ($TAM_u$) are not same.

**4. The Rectangle Packing Problem**

The concept of using rectangles for core test representation has been used before in [4,6,8]. Consider a SOC having N cores and let $R_i$ be the set of rectangles for core i, $1 \leq i \leq N$. Generalized version of rectangle packing problem PROBLEM-$_{RP}$ is as follows: select a rectangle R from $R_i$ for each set $R_i$, $1 \leq i \leq N$ and pack the selected rectangles in a bin of fixed height and unbounded width such that no two rectangle overlap and the width to which the bin is filled is minimized. Unlike in [6], each rectangle selected is not allowed to be split vertically in our rectangle packing.

| TAM size | TAM utilized ($TAM_u$) | Longest Scan chain |
|---|---|---|
| 50-64 | 47 | 521 |
| 48-49 | 39 | 1021 |
| 32-47 | 24 | 1042 |
| 24-31 | 16 | 1563 |
| 20-23 | 12 | 2084 |
| 16-19 | 10 | 2605 |
| 14-15 | 8 | 3126 |
| 12-13 | 7 | 3647 |
| 10-11 | 6 | 4689 |
| 8-9 | 5 | 5729 |
| 6-7 | 4 | 7809 |
| 4-5 | 3 | 11969 |
| 2-3 | 2 | 23789 |
| 1 | 1 | 24278 |

Table 1: result of Wrapper_Design for core 6 of p93791 [11]

In this paper, the wrapper/TAM co-optimization problem PROBLEM-$_{OPT}$ that we consider is as follows: determine the TAM width to be assigned and design a wrapper for each core and schedule the tests for the SOC in such a way that minimizes the total testing time and the total number of TAM wires utilized at any moment does not exceed total TAM width when a set of parameters for each core is given. The set of parameters for each core includes the number of primary I/Os, test patterns, scan chains and scan chain lengths.

---

Data structure test_schedule

1. width[i]    //TAM width assigned to core i
2. finish[i]   //end time of core i
3. scheduled[i] //boolean indicates core i is scheduled
4. start[i]    //begin time of core i
5. complete[i] //boolean indicates test for core i has finished
6. peak_tam[i] //equals to MAX_$TAM_u$

---

Figure 2: Data structure for the test schedule

We solve the PROBLEM-$_{OPT}$ by generalized version of rectangle packing or two-dimensional packing Problem-$_{RP}$. We use the Wrapper_Design algorithm to obtain the different test times for each core for varying values of TAM width. A set of rectangles for a core can now be constructed, such that the height of each rectangle corresponds to a different TAM width and the width of the rectangle represents the core test application time for this value of TAM width.

PROBLEM-$_{RP}$ relates to PROBLEM-$_{OPT}$ as follows: The height of the rectangle selected for a core corresponds to the TAM width assigned to the core, while the rectangle width corresponds to its testing time. The height of the bin corresponds to the total SOC TAM width, and the width to which the bin is ultimately filled corresponds to the system testing time that is to be minimized. The unfilled area of the bin corresponds to the idle time on TAM wires during test. Furthermore, the distance between the left edge of each rectangle and the left edge of the bin corresponds to the begin time of each core test.

Our approach emphasizes on both testing time of a core and the TAM width required to achieve that testing time by considering the diagonal length of rectangles. Consider three rectangles R[1] = {H=32, W=7.1, DL=32.78}, R[2] = {H=16, W=13.8, DL=21.13}, R[3] = {H=32, W=5.4, DL=32.45) where W,H, DL denotes width, height and diagonal length of the rectangles respectively. Here if we take into account testing time(W), then we should pack R[2] first, followed by R[1] and R[3]. We found that this does not produce best result in rectangle packing. But when we consider diagonal lengths, we pack R[1], R[3], R[2] in sequence, and get the result that is extremely efficient.

---

Algorithm Test_Scheduling ($W_{max}$, Core C[1...NC])

{

1. For each core C[i], construct a set of rectangles taking $TAM_u$ as rectangle height and its corresponding testing time as rectangle width such that $TAM_u <= W_{max}$

2. Find the smallest($T_{min}$) among the testing time corresponding to MAX_$TAM_u$ of all cores

3. For each core C[i], divide the width T[i] of all rectangles constructed in line 1 with $T_{min}$.

4. For each core C[i], calculate Diagonal Length DL[i] = $\sqrt{(W[i])^2 + (T[i]^2)}$ where W[i] denotes MAX_$TAM_u$ and T[i] denotes corresponding reduced testing time.

5. Sort the Cores in descending order of diagonal length calculated in line 4 and keep in list INITIAL[NC]

6. Next_Schedule_Time = 0
   current_Time = 0;
   Wavail = $W_{max}$;   // TAM available
   Idle_Flag=False;
// peak_tam[c] is equal to MAX_$TAM_u$ of core c
//   PENDING is a queue.

7. While (INITIAL and PENDING not Empty)
   {
   8 If (Wavail > 0 and Idle_Flag=False )
        {
      9. If (INITIAL is not empty)
         {
         c=delete(INITIAL);
         If ( Wavai ≥ peak_tam[c])
            Update(c,peak_tam(c));
         Else If(Possible_TAM ≥ 0.5*peak_tam[c])
            Update(c, Possible_TAM);
         Else
            add(PENDING,c);
         if(peak_tam[PENDING[front]] ≤ Wavail)
            Update(PENDING[front],
                peak_tam[PENDING[front]]);
            delete(PENDING) ;
         }
      10. Else //if INITIAL is empty
         {
         If(peak_tam[PENDING[front]] ≤ Wavail)
            Update(PENDING[front],
                peak_tam[PENDING[front]]);
            delete(PENDING) ;
         Else
            Idle_Flag=True;
         }
            }

11. Else  //TAM available < 0 or idle
   {
   Calculate Next_Schedule_Time = Finish[i],
      Such that Finish[i]> This_Time and Finish[i] is minimum;
   Set This_Time=Next_Schedule_Time;

   12. For every Core i, such that finish[i] = This_Time

         Wavail = Wavail + Width[i];
         13. Set Complete[i] = TRUE;
   Idle_Flag=False;
   }
} //end of while
   return  test_schedule;

}

---

Figure3: proposed Test scheduling algorithm with TAM optimization

Procedure update( i , w)
---
1. Let i be the core to be updated in the test schedule
2. Start[i]=Current_Time;
3. Set scheduled[i] = TRUE;
4. finish[i] = Current_Time + $T_i(w)$;
5. width[i]=w;
6. Wavail=Wavail- w;

---

Figure 4:Data structure for the update algorithm

**5. Proposed Test Scheduling**

**Rectangle construction:** In our proposed test scheduling algorithm (figure 3), after getting the result of Wrapper_Design, for each core, we construct a set of rectangles taking $TAM_u$ as rectangle height and its corresponding testing time as rectangle width such that $TAM_u \leq W_{max}$ (figure 5) rather than constructing the collection of Pareto-optimal rectangles like [5]. MAX_$TAM_u$ is the largest among the $TAM_u$ values satisfying the above constraint. In figure 5, MAX_$TAM_u$=24 and $W_{max=32}$. For combinational core, MAX_$TAM_u$ is always equal to $W_{max}$. Note that, In case of TAM wire assignment to that particular scheduling of p93791 (figure 5), TAM wires that are to be assigned to core 6 must be selected from values 24,16,12,10,8-1 depending on TAM width available

**Diagonal length calculation**: In line 2, we find the smallest($T_{min}$) among the testing time corresponding to MAX_$TAM_u$ for all cores. In line 3,for each core we divide width(testing time) of all constructed rectangles ( line 3) with $T_{min}$. Then in line 4,for each core we calculate the diagonal length of the rectangle where rectangle height W[i] =MAX_TAM$_u$ and rectangle width T[i] is reduced testing time corresponding to MAX_TAM$_u$. We then sort the cores in descending order of diagonal length calculated in line 4.

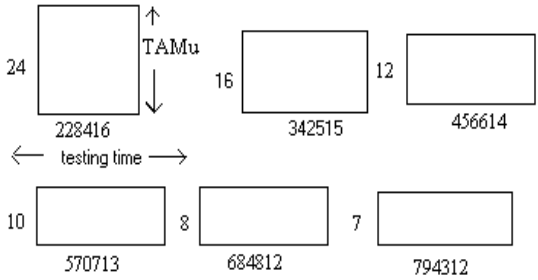

Figure 5: example of some rectangles for core 6 of SOC p93791(figure drawn not to scale) when W$_{max=32}$

**TAM assignment:** While executing the main **While** loop(line 7),if there are Wavail TAM wires available for assignment and list INITIAL is not empty, we select a core c from the list in sorted order. If TAM available at that moment ( Wavail ) is greater than or equal to peak_tam[c],we schedule the tests of that core and assign TAM wires to c equal to peak_tam[c].Note that ,peak_tam[c] is equal to MAX_TAM$_u$ of core c. If Wavail is less than peak_tam[c],it tries to find a TAM$_u$ value such that TAM$_u$ $\leq$ Wavail and TAM$_u$ greater than half of peak_tam[c]. If it fails to assign TAM wires to c satisfying these conditions, it add the core c into queue PENDING.It then deletes a core p from the queue PENDING for scheduling only if Wavail is greater than or equal to peak_tam[p].

If list INITIAL is empty, the algorithm deletes the core c at the front of queue PENDING only if Wavail $\geq$ peak_tam[c].Otherwise it waits until sufficient TAM wires become available. If Wavail>0 and INITIAL is empty, these Wavail wires are declared idle and Idle_Flag is set if Wavail cannot satisfy the condition Wavail $\geq$ peak_tam[c] where c is the core at the front of queue PENDING .

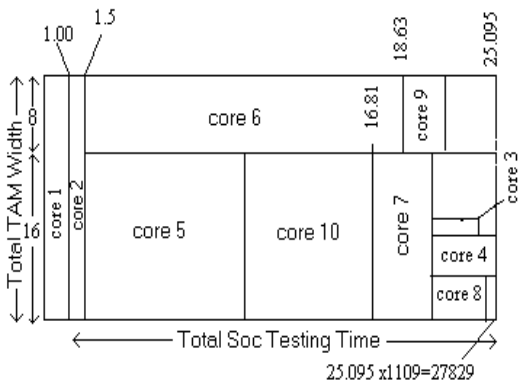

Figure 6:Test scheduling for d695 using our algorithm (T$_{min}$=1109 and TAM width=24)

If there are Wavail idle wires or Wavail=0,the execution proceeds to line 12 where the process of updating This_Time to Next_Schedule_Time and Wavail is begun .Line 13 increases Wavail by the width of all cores ending at the new value of This_Time and Line 13 sets complete[i] to true for all cores whose test has completed at This_Time.

**6. Experimental results**

In this section, we present experimental results for one example SOC: d695. This SOC is a part of the ITC'02 SOC benchmarking initiative[11].In our algorithm we considered TAM wire sharing as test conflict. The results for SOC d695 are given in Table 2. In this Table we compare the testing times obtained using our proposed approach and previous approaches of wrapper/TAM co-optimization for a given TAM width. Note that none of the previous approaches consider more test conflicts than TAM wire sharing.

**7. Conclusion**

In this paper, we have presented a new technique based on rectangle packing for Wrapper/TAM co-optimization and test scheduling .We have emphasized on both time and TAM width by considering diagonal lengths. The experimental results show the efficiency of our algorithm.

| TAM Width | [5] | [6] | [7] | Proposed |
|---|---|---|---|---|
| 64 | 12941 | 11604 | 12941 | 14914 |
| 56 | 13207 | 13415 | 12941 | 16242 |
| 48 | 16975 | 15698 | 15300 | 16317 |
| 40 | 17901 | 18459 | 18448 | 20207 |
| 32 | 21566 | 23021 | 22268 | 20402 |
| 24 | 28292 | 30317 | 30032 | 27829 |
| 16 | 42568 | 43723 | 42644 | 39572 |

Table 2:Experimental result for d695

**8.References**